%% file: master.tex
\documentclass[runningheads]{llncs}
\usepackage{graphicx}

\usepackage{amsmath,amssymb,amsfonts}
\usepackage{xcolor}
\usepackage{interval}
\usepackage{placeins}
\usepackage[labelformat=simple]{subcaption}

\usepackage{sansmath}
\usepackage{uppaal}
\usepackage[inkscapelatex=false]{svg}
\usepackage{acro}
\usepackage[inline]{enumitem}
\usepackage{comment}
\usepackage{wrapfig}
\usepackage{xfrac}
\usepackage{booktabs}
\usepackage{multirow}
\usepackage{tikz}

\usepackage{algorithm,algpseudocode}
\algnewcommand{\To}{\textbf{To }}
\algnewcommand{\Input}{\item[\textbf{Input:}]}%
\algnewcommand{\Output}{\item[\textbf{Output:}]}%

\usetikzlibrary{arrows,shapes,positioning}
\usetikzlibrary{calc,decorations.markings}

\usepackage{hyperref}
\input{macros}

\input{acronyms}

\begin{document}
\title{Shielded Reinforcement Learning\\ for Hybrid Systems
}

%
%
\author{Asger Horn Brorholt \and
{Peter Gjøl Jensen} \and
{Kim Guldstrand Larsen} \and
{Florian~Lorber} \and
{Christian~Schilling}}
\authorrunning{Brorholt, Jensen, Larsen, Lorber and Schilling}
%
\institute{Department of Computer Science, Aalborg University, Aalborg, Denmark\\
\email{\{asgerhb,pgj,kgl,florber,christianms\}@cs.aau.dk}}
\maketitle              
\begin{abstract}
Safe and optimal controller synthesis for switched-controlled hybrid systems, which combine differential equations and discrete changes of the system's state, is known to be intricately hard. Reinforcement learning has been leveraged to construct near-optimal controllers, but their behavior is not guaranteed to be safe, even when it is encouraged by reward engineering. One way of imposing safety to a learned controller is to use a \emph{shield}, which is correct by design. However, obtaining a shield for non-linear and hybrid environments is itself intractable. In this paper, we propose the construction of a shield using the so-called \emph{barbaric method}, where an approximate finite representation of an underlying partition-based two-player safety game is extracted via systematically picked samples of the true transition function. While hard safety guarantees are out of reach, we experimentally demonstrate strong statistical safety guarantees with a prototype implementation and \uppaalstratego. Furthermore, we study the impact of the synthesized shield when applied as either a pre-shield (applied before learning a controller) or a post-shield (only applied after learning a controller). We experimentally demonstrate superiority of the pre-shielding approach. We apply our technique on a range of case studies, including two industrial examples, and further study post-optimization of the post-shielding approach.

\end{abstract}

\input{intro}

\input{EMDP.tex}

\input{computingreachability}

\input{shield}

\input{results}

\input{conclusion}

\section*{Acknowledgments}

This research was partly supported by DIREC - Digital Research Centre Denmark and the Villum Investigator Grant S4OS - Scalable analysis and Synthesis of Safe,  Secure and Optimal Strategies for Cyber-Physical Systems.

\FloatBarrier

\bibliographystyle{splncs04}
\bibliography{sources.bib}{}

\end{document}

%% file: macros.tex
\newcounter{examplectr}
\newcommand{\exampleend}{\hfill$\triangleleft$}

\newcommand{\uppaalstratego}{\textsc{Uppaal Stratego}\xspace}

\newcommand{\juliareach}{\textsc{Julia\-Reach}\xspace}

\newcommand{\statespace}{\mathcal{S}}
\newcommand{\partitioning}{\mathcal{A}}

\newcommand{\granularity}{\gamma}

\newcommand{\state}{s}
\newcommand{\partition}{\mu}
\newcommand{\action}{a}
\newcommand{\transitionfunction}{T}
\newcommand{\transitionsystem}{{\cal T}^{{\cal A}}_{{\cal M}}}
\newcommand{\approxts}{\widehat{\cal T}^{{\cal A}}_{{\cal M}}}
\newcommand{\costfunction}{C}

\newcommand{\samplesperaxis}{N}

\newcommand{\curlyg}{\mathcal{G}}

\newcommand{\curlym}{\mathcal{M}}

\newcommand{\act}{\textit{Act}}
\newcommand{\Idelta}{\mathcal{I}_\granularity}

\newcommand{\integers}{\mathbb{Z}}
\newcommand{\realnumber}{\mathbb{R}}

\newcommand{\suchthat}{~.~}
\newcommand{\onehalf}{{1 \over 2}}

\newcommand{\abs}[1]{\left | #1 \right |}

\newcommand{\states}{\mathcal{S}}

\newcommand{\mdp}{\mathcal{M}}

\newcommand{\run}{\pi}

\newcommand{\goal}{\mathcal{G}}

\newcommand{\expected}{\ensuremath{\mathbb{E}}}

\definecolor{turquoise}{HTML}{1ABC9C}
\definecolor{emerald}{HTML}{2ECC71}
\definecolor{peterriver}{HTML}{3498DB}
\definecolor{amethyst}{HTML}{9B59B6}
\definecolor{wetasphalt}{HTML}{34495E}

\definecolor{greensea}{HTML}{16A085}
\definecolor{nephritis}{HTML}{27AE60}
\definecolor{belizehole}{HTML}{2980B9}
\definecolor{wisteria}{HTML}{8E44AD}
\definecolor{midnightblue}{HTML}{2C3E50}

\definecolor{sunflower}{HTML}{F1C40F}
\definecolor{carrot}{HTML}{E67E22}
\definecolor{alizarin}{HTML}{E74C3C}
\definecolor{clouds}{HTML}{ECF0F1}
\definecolor{concrete}{HTML}{95A5A6}

\definecolor{orange}{HTML}{F39C12}
\definecolor{pumpkin}{HTML}{D35400}
\definecolor{pomegranate}{HTML}{C0392B}
\definecolor{silver}{HTML}{BDC3C7}
\definecolor{asbestos}{HTML}{7F8C8D}

\newcommand{\todo}[1] {

\noindent\textcolor{pumpkin}{\texttt{\hspace{-3em}todo:}} {\small\textsf{#1}}

}

\newcommand{\svg}[3] {
\begin{figure}[ht]
    \centering
    \includesvg[#2]{graphics/#1.svg}
    \caption{#3}\label{fig:#1}
\end{figure}
}

\newcommand{\triplesvgspace}[9]{
\begin{figure}[#9]
    \centering
    \begin{subfigure}{.48\textwidth}
        \centering
        \includesvg[width=.95\linewidth]{graphics/#1.svg}
        \caption{#2}
        \label{fig:#1}
    \end{subfigure}
    \hfill
    \begin{subfigure}{.51\textwidth}
        \begin{subfigure}{1.0\textwidth}
            \centering
            \includesvg[width=.95\linewidth]{graphics/#3.svg}
            \caption{#4}
            \label{fig:#3}
        \end{subfigure}
        \begin{subfigure}{1.0\textwidth}
            \centering
            \includesvg[width=.95\linewidth]{graphics/#5.svg}
            \caption{#6}
            \label{fig:#5}
        \end{subfigure}
    \end{subfigure}
    \caption{#8}
    \label{fig:#7}
    \vspace{-10pt}
\end{figure}
}

%% file: acronyms.tex
\DeclareAcronym{rl}{
  short = RL,
  long  = reinforcement learning,
  short-indefinite = an,
  long-indefinite = a
}

%% file: intro.tex
\section{Introduction}\label{sec:introduction}

Digital controllers are key components of cyber-physical systems.
Unfortunately, the algorithmic construction of controllers is intricate for any but the simplest systems~\cite{lewis2012optimal,doyle2013feedback}.
This motivates the usage of \ac{rl}, which is a powerful machine-learning method applicable to systems with complex and stochastic dynamics~\cite{BusoniuBTKP18}.

However, while controllers obtained from \ac{rl} provide near-optimal average-case performance, they do not provide guarantees about worst-case performance, which limits their application in many relevant but safety-critical domains, ranging from power converters to traffic control~\cite{VlachogiannisH04,NoaeenNGCAABF22}.
A typical way to tackle this challenge is to integrate safety into the optimization objective via \textit{reward shaping} during the learning phase, which punishes unsafe behavior~\cite{10.5555/2789272.2886795}.
This will make the controller more robust to a certain degree, but safety violations will still be possible, and the integration of safety into the optimization objective can reduce the performance, thus yielding a controller that is neither safe nor optimal.

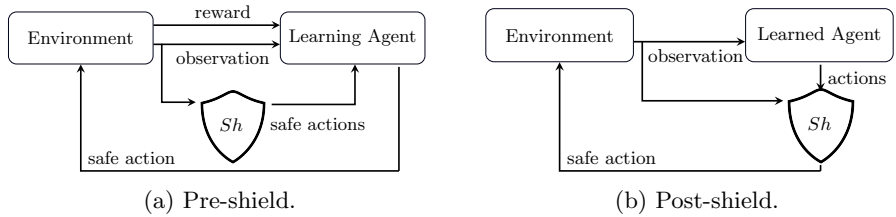
\begin{figure}[t]
    \centering
    \begin{subfigure}{.48\linewidth}
        \centering
        \resizebox{.99\linewidth}{!}{
            \input{graphics/preshielding}
        }
        \caption{Pre-shield.}
        \label{fig:pre-shield}
    \end{subfigure}
    \hfill
    \begin{subfigure}{.48\linewidth}
        \centering
        \resizebox{.99\linewidth}{!}{
            \input{graphics/postshielding}
        }
        \caption{Post-shield.}
        \label{fig:pst-shield}
    \end{subfigure}
    \caption{Pre- and post-shielding in a reinforcement-learning setting.}
    \label{fig:shieldingTypes}
\end{figure}

A principled approach to obtain worst-case guarantees is to use a \emph{shield} that restricts the available actions~\cite{DBLP:conf/tacas/BloemKKW15}.
This makes it possible to construct correct-by-design and yet near-optimal controllers.
Fig.~\ref{fig:shieldingTypes} depicts two ways of shielding RL agents: {\em pre-} and {\em post-shielding}.
Pre-shielding is already applied during the learning phase, and the learning agent receives only safe actions to choose from.
Post-shielding is only applied during deployment, where the trained agent is monitored and, if necessary, corrected.
Such interventions to ensure safety interfere with the learned policy of the agent, potentially causing a loss in optimality.

In a nutshell, the algorithm to obtain a shield works as follows.
First we compute a finite partitioning of the state space and approximate the transitions between the partitions.
This results in a two-player safety game, and upon solving it, we obtain a strategy that represents the most permissive shield.

Cyber-physical systems exhibit behavior that is both continuous (the environment) and discrete (the control, and possibly the environment too).
We are particularly interested in a class of systems we refer to as \emph{hybrid Markov decision processes} (HMDPs).
In short, these are control systems where the controller can choose an action in a periodic manner, to which the environment chooses a stochastic continuous trajectory modeled by a stochastic hybrid automaton~\cite{DBLP:journals/corr/abs-1208-3856}.
While HMDPs represent many real-world systems, they are a rich extension of hybrid automata, and thus their algorithmic analysis is intractable even under serious restrictions~\cite{HenzingerKPV98}.
These complexity barriers unfortunately also carry over to the above problem of constructing a shield.

\smallskip

In this paper, we propose a new practical technique to automatically and robustly synthesize a shield for HMDPs.
The intractability in the shield-synthesis algorithm is due to the rigorous computation of the transition relation in the abstract transition system, since that computation reduces to the (undecidable) reachability problem.
Our key to get around this limitation is to approximate the transition relation through systematic sampling, in a way that is akin to the \textit{barbaric method} (a term credited to Oded Maler~\cite{KapinskiKMS03,Donze10}).

We combine our technique with the tool \uppaalstratego to learn a shielded near-optimal controller, which we evaluate in a series of experiments on several models, including two real-world cases.
In our experiments we also find that pre-shielding outperforms post-shielding.
While the shield obtained through our technique is not guaranteed to be safe in general due to the approximation, we demonstrate that the controllers we obtain are statistically safe, and that a moderate number of samples is sufficient in practice.

\vspace{0.2cm}
\emph{Related work.}
Enforcing safety during RL by limiting the choices available to the agent is a known concept, which is for instance applied in the tool \uppaalstratego~\cite{stratego}.
The term ``shielding'' was coined by Bloem et al.~\cite{DBLP:conf/tacas/BloemKKW15}, who introduced special conditions on the enforcer like \textit{shields with minimal interference} and \textit{k-stabilizing shields} and later demonstrated shielding for RL agents~\cite{10.5555/3504035.3504361}, where they correct potentially unsafe actions chosen by the RL agent.
Jansen et al.~\cite{jansen2020safe} introduced shielding in the context of RL for probabilistic systems.
A concept similar to shielding has also been proposed for safe model predictive control~\cite{BastaniL21,WabersichZ21}.
Carr et al.~\cite{Carr0JT23} show how to shield partially observable environments.
In a related spirit, Maderbacher et al.\ start from a safe policy and switch to a learned policy if safe at run time~\cite{MaderbacherSBBNK23}.

(Pre-)Shielding requires a model of the environment in order to provide safety guarantees during learning.
Orthogonal to shielding, several model-free approaches explore an \ac{rl} environment in a \emph{safer} way, but without any guarantees.
Several works are based on barrier certificates and adversarial examples~\cite{ChengOMB19,LuoM21} or Lyapunov functions~\cite{HasanbeigAK20}.
Similarly, Berkenkamp et al.\ describe a method to provide a safe policy with high probability~\cite{BerkenkampTS017}.
Chow et al.\ consider a relaxed version of safety based on expected cumulative cost~\cite{ChowNDG18}.
In contrast to these model-free approaches, we assume a model of the environment, which allows us to safely synthesize a shield just from simulations before the learning phase.
We believe that the assumption of a model, typically derived from first principles, is realistic, given that our formalism allows for probabilistic modeling of uncertainties.
To the best of our knowledge, none of the above works can be used in practice for safe \ac{rl} in the complex class of HMDPs.

Larsen et al.~\cite{10.1007/978-3-030-23703-5_6} used a set-based Euler method to overapproximate reachability for continuous systems. This method was used to obtain a safety strategy and a safe near-optimal controller. Contrary to that work, we apply both pre- and post-shielding, and our method is applicable to more general hybrid systems.
We employ state-space partitioning, which is common for control synthesis~\cite{MajumdarOS20} and reachability analysis~\cite{KlischatA20} and is also used in recent work on learning a safe controller for discrete stochastic systems in a teacher-learner framework~\cite{ZikelicLHC23}. 
Contemporary work by Badings et al.~\cite{badings_robust_2023} also uses a finite state-space abstraction along with sample-based reachability estimation, to compute a reach-avoid controller. 
The method assumes linear dynamical systems with stochastic disturbances, to obtain upper and lower bounds on transition probabilities.
In contrast, our method supports a very general hybrid simulation model, and provides a safety shield, which allows for further optimization of secondary objectives.

A special case of the HMDPs we consider is the class of stochastic hybrid systems (SHSs).
Existing reachability approaches are based on state-space partitioning~\cite{AbateAPLS07,ShmarovZ15}, which we also employ in this work, or have a statistical angle~\cite{Bujorianu12}.
We are not aware of any works that extended SHSs to HMDPs.

\vspace{0.2cm}
\emph{Outline.} The remainder of the paper is structured as follows.
In Section~\ref{sec:EMDP} we present the formalism we use.
In Section~\ref{sec:computingreachability} we present our synthesis method to obtain a safety strategy and explain how this strategy can be integrated into a shield.
We demonstrate the performance of our pre- and post-shields in several cases in Section~\ref{sec:experiments}.
Finally we conclude the paper in Section~\ref{sec:conc}.

%% file: graphics/preshielding.tex
\begin{tikzpicture}
\definecolor{myblack}{cmyk}{.67,.33,0,.99}
\tikzstyle{box} = [rectangle, rounded corners, minimum width=2.5cm, minimum height=0.9cm, text centered, draw=myblack]
\tikzstyle{arrow} = [thick,->,>=stealth]
\tikzstyle{point} = [coordinate]
\definecolor{dgreen}{cmyk}{60, 0,100,0}

\node (env) [box] {Environment};
\node[inner sep=1pt, right of=env, xshift=2.0cm, yshift=-1.2cm] (shield) {{{\includegraphics[bb= 15 10 800 780,scale=0.07]{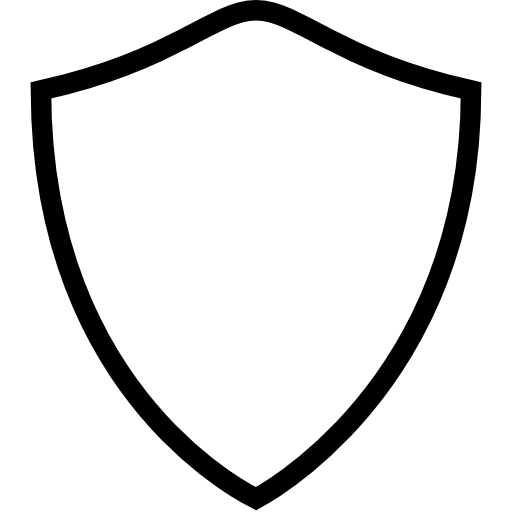}}}};
\node[text width=1cm] at (2.9,-1.5) {$Sh$};
\node (la) [box, right of=shield, xshift=0.7cm,yshift=1.2cm] {Learning Agent};


\draw [arrow, thick] (env.10) -- node [above]{~~reward} (la.170);
\draw [arrow, thick] (env.-5) -- node [anchor=north]{~~observation} (la.185);

\draw [arrow, thick] (1.4, -0.1) |- node [anchor=south]{} (shield.175);

\draw [arrow, thick] (3.3,-1.15) -| node [below left = 0.1 and -0.3 ]{safe actions} (la.275);

\draw [arrow, thick] (5.5,-2.3) -| node[pos=0.42, above] {safe action} (env);


\draw[thick] (5.5, -0.5) --  (5.50,-2.3);

\end{tikzpicture}

%% file: graphics/postshielding.tex
\begin{tikzpicture}
\definecolor{myblack}{cmyk}{.67,.33,0,.99}
\tikzstyle{box} = [rectangle, rounded corners, minimum width=2.5cm, minimum height=0.9cm, text centered, draw=myblack]
\tikzstyle{arrow} = [thick,->,>=stealth]
\tikzstyle{point} = [coordinate]
\definecolor{dgreen}{cmyk}{60, 0,100,0}

\node (env) [box] {Environment};
\node (la) [box, right of=env, xshift=3.4cm] {Learned Agent};


\node[inner sep=1pt, right of=env, xshift=3.8cm, yshift=-1.2cm] (shield) {{{\includegraphics[bb= 15 10 800 780,scale=0.07]{graphics/shield-icon-blank}}}};

\draw [arrow, thick] (env.-5) -- node [anchor=north]{~~observation} (la.185);

\draw [arrow, thick] (1.4, -0.1) |- node [anchor=south]{} (shield.175);

\draw [arrow, thick] (4.42,-2.3) -| node[pos=0.4, above] {safe action} (env);

\draw [arrow, thick] (4.42, -0.5) -- node [anchor=west]{actions} (4.42, -0.95);
\node[text width=1cm] at (4.7,-1.5) {$Sh$};
\draw[thick] (4.42, -2.2) --  (4.42,-2.3);

\end{tikzpicture}

%% file: EMDP.tex
\section{Euclidian and Hybrid Markov Decision Processes}
\label{sec:EMDP}

In this section we introduce the system class we study in this paper: hybrid Markov decision processes (HMDPs).
They combine Euclidean Markov decision processes and stochastic hybrid automata, which we introduce next.
HMDPs model complex systems with continuous, discrete and stochastic dynamics.

\subsubsection{Euclidean Markov Decision Processes}
A Euclidean Markov decision process (EMDP)~\cite{DBLP:conf/atva/JaegerJLLST19,randomwalk} is a continuous-space extension of a Markov decision process (MDP).
We recall its definition below.

\begin{definition}[Euclidean Markov decision process]
A \emph{Euclidean Markov decision process} of dimension $k$ is a tuple $\mdp=(\statespace, \state_0, \act, \transitionfunction, \costfunction, \goal)$ where
\begin{itemize}
    \item $\statespace \subseteq \realnumber^k$ is a bounded and closed part of $k$-dimensional Euclidean space,
    \item $\state_0 \in \statespace$ is the initial state,
    \item $\act$ is the finite set of actions,
    \item $\transitionfunction: \statespace \times \act \rightarrow (\statespace \rightarrow \realnumber_{\geq0} )$ maps each state-action pair $(s,a)$ to a probability density function over $\statespace$, i.e., we have $\int_{s'\in \statespace} \transitionfunction(s,a)(s') ds'=1$,
    \item $\costfunction: \statespace \times \act \times \statespace \rightarrow \realnumber$ is the cost function, and
    \item ${\cal G}\subseteq \statespace$ is the set of goal states.
\end{itemize}
\end{definition}


\begin{wrapfigure}{r}{0.37\textwidth}
    \vspace{-20pt}
    \centering
    \includesvg[width=\linewidth]{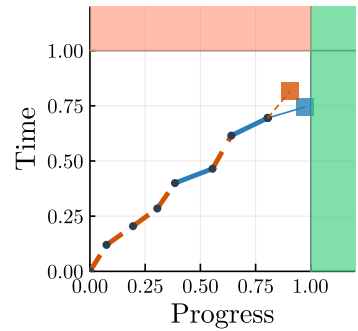}
    \caption{A random walk with action sequence \emph{slow, slow, slow, slow, fast, slow, fast}.}
    \vspace{-15pt}
    \label{fig:RandomWalk}
\end{wrapfigure}
    \refstepcounter{examplectr}
    \paragraph{Example~{\theexamplectr} \label{ex:random_walk} (Random walk).}
    Fig.~\ref{fig:RandomWalk} illustrates an EMDP of a (semi-)random walk on the state space $\statespace=[0,x_{max}]\times[0,t_{max}]$ (one-dimensional space plus time).
    The goal is to cross the $x=1$ finishing line before $t=1$.
    Two movement actions are available:
    fast and expensive (blue), or slow and cheap (brown).
    Both actions have uncertainty about the distance traveled and time taken.
    Given a state $(x,t)$ and an action $a\in\{\mathit{slow}, \mathit{fast}\}$, the next-state density function $\transitionfunction((x,t),a)$ is a uniform distribution over the successor-state set
    $(x+d_x(a) \pm \epsilon) \times (t+d_t(a) \pm \epsilon)$,
    where $d_x(a)$ and $d_t(a)$ respectively represent the direction of movement in space and time given action $a$, while $\epsilon$ models the uncertainty.\exampleend
\medskip

A run $\run$ of an EMDP is an alternating sequence $s_0a_0s_1a_1\dots$ of states and actions such that $T(s_i,a_i)(s_{i+1})>0$ for all $i\geq0$.
A (memoryless) strategy for an EMDP is a function $\sigma:\states\rightarrow(\act\rightarrow[0,1])$, mapping a state to a probability distribution over $\act$.
Given a strategy $\sigma$, the expected cost of reaching a goal state is defined as the solution to a Volterra integral equation as follows:

\begin{definition}[Expected cost of a strategy]
\label{def:expcost}
  Let $\mdp=(\statespace, \state_0, \act, \transitionfunction, \costfunction, \goal)$ be an EMDP and $\sigma$ be a strategy.
  If a state $s$ can reach the goal set $\goal$, the \emph{expected cost}
  is the solution to the following recursive equation:
  \begin{equation*}
      \expected_\sigma^\mdp(s) = \begin{cases}
            0 & \text{if } s \in \goal \\
            \displaystyle\sum_{a\in\act}\sigma(s)(a)\cdot\int_{s'\in\states}
            T(s,a)(s')\cdot\big( \costfunction(s,a,s') +
            \expected_\sigma^\mdp(s') \big) \, ds' & \text{if } s \notin \goal
            \end{cases}
  \end{equation*}
\end{definition}

A strategy $\sigma^*$ is optimal if it minimizes $\expected_\sigma^\mdp(s_0)$.
We note that there exists an optimal strategy which is deterministic.

\subsubsection{Stochastic Hybrid Systems}
In an EMDP, the environment responds instantaneously to an action  proposed by the agent according to the next-state density function $\transitionfunction$. In a more refined view, the agent proposes actions with some period $P$, and the response of the environment is  a stochastic, time-bounded trajectory (bounded by the period $P$) over the state space. For this response, we use a stochastic hybrid system (SHS)~\cite{DBLP:journals/corr/abs-1208-3856,DBLP:conf/formats/Larsen12}, which allows the environment to interleave continuous evolution and discrete jumps.

\begin{figure}[tb]
    \centering
    \begin{subfigure}[b]{0.49\textwidth}
        \centering
        \includegraphics[width=0.96\textwidth]{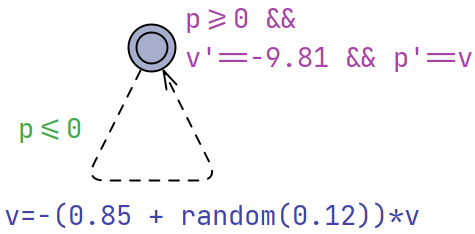}
        \caption{SHA for the bouncing ball.}
    \end{subfigure}
    \hfill
    \begin{subfigure}[b]{0.49\textwidth}
        \centering
        \includesvg[width=\textwidth]{graphics/BallBounceDensity.svg}
        \caption{State density after one bounce.}
    \end{subfigure}
    \caption{An SHA for the bouncing ball and a visualization after one bounce.}
    \label{fig:BBBehaviour}
\end{figure}

\begin{definition}[Stochastic hybrid system]
A \emph{stochastic hybrid system} of dimension $k$ is a tuple  ${\cal H}=(\statespace,F,\mu,\eta)$ where
\begin{itemize}
     \item $\statespace \subseteq \realnumber^k$ is a bounded and closed part of $k$-dimensional Euclidean space,
    \item $F: \realnumber_{\geq 0}\times \statespace \rightarrow \statespace$ is a flow function  describing the evolution of the continuous state with respect to time, typically represented by differential equations,
     \item $\mu: \statespace \rightarrow (\realnumber_{\geq0}\rightarrow\realnumber_{\geq0})$ maps each state $s$ to a delay density function $\mu(s)$ determining the time point for the next discrete jump, and
    \item $\eta: \statespace\rightarrow (\statespace\rightarrow \realnumber_{\geq0})$ maps each state $s$ to a density function $\eta(s)$ determining the next state.
\end{itemize}
\end{definition}

\refstepcounter{examplectr}
\paragraph{Example~{\theexamplectr} (Bouncing ball).}
To represent an SHS, we use a stochastic hybrid automaton (SHA)~\cite{DBLP:journals/corr/abs-1208-3856}, which we only introduce informally here.
Fig.~\ref{fig:BBBehaviour}(a) shows an SHA of a bouncing ball, which we use as a running example.
Here the state of the ball is given by a pair $(p,v)$ of continuous variables, where $p\in\realnumber_{\geq0}$ represents the current height (position) and $v\in\realnumber$ represents the current velocity of the ball.
Initially (not visible in the figure) the value of $v$ is zero while $p$ is picked randomly in $\interval{7.0}{10.0}$.
The behavior of the ball is defined by two differential equations: $v'=-9.81 m/s^2$ describing the velocity of a falling object and $p'=v$ stating that the rate of change of the height is the current velocity.
The invariant $p\geq 0$ expresses that the height is always nonnegative.
The single transition of the automaton triggers when $p\leq0$, i.e., when the ball hits the ground.
In this case the velocity reverts direction and is subject to a random dampening effect (here ``\uppUpdate{\texttt{random(0.12)}}'' draws a random number from $[0, 0.12]$ uniformly).
The state density after one bounce is illustrated in Fig.~\ref{fig:BBBehaviour}(b).
The SHA induces the following SHS, where $\delta$ denotes the Dirac delta distribution:
\begin{itemize}
\item $\statespace= [0, 10] \times [-14, 14]$,
\item $F((p,v),t) = ((-9.81/2)t^2 + vt +p, -9.81t +v)$
\item $\mu((p,v)) = 
  \delta\big( (v + \sqrt{v^2 +2\cdot 9.81 \cdot p})/9.81 \big)$
\item $\eta((p,v)) = (p, v\cdot{\cal U}_{[-0.97,-0.85]})$, with uniform distribution ${\cal U}_{[l,u]}$ over $[l,u]$.\exampleend
\end{itemize}

\medskip

A timed run $\rho$ of an SHS ${\cal H}$ with $n$ jumps from an initial state density $\iota$ is a sequence $\rho=s_0 s'_0 t_0 s_1 s'_1 t_1 s_2 s'_2 \dots t_{n-1} s_n s'_n$ respecting the constraints of ${\cal H}$, where each $t_i\in\realnumber_{\geq0}$. The total duration of $\rho$ is $\sum_{i=0}^{n-1} t_i$, and the density of $\rho$ is $\iota(s_0)\cdot\prod_{i=0}^{n-1}\mu(s'_i)(t_i)\cdot \eta(s_{i+1})(s'_{i+1})$.

Given an initial state density $\iota$ and a time bound $T$, we denote by $\Delta_{{\cal H},\iota}^T$  the density function on $\statespace$ determining the state after a total delay of $T$, when starting in a state given by $\iota$. The following recursive equation defines $\Delta_{{\cal H},\iota}^T$:\footnote{For SHS with an upper bound on the number of discrete jumps up to a given time bound $T$, the equation is well-defined.}
\[
\Delta_{{\cal H},\iota}^T(s') = \begin{cases}
        \iota(s') & \text{if } T = 0 \\
        \displaystyle\int_s \iota(s)\cdot \int_{t\leq T}
            \mu(s)(t)\cdot \Delta_{{\cal H},\eta(F(t,s))}^{T-t}(s') \, dt \, ds & \text{if } T>0
                                \end{cases}
\]

For $T=0$, the density of reaching $s'$ is given by the initial state density function $\iota$. 
For $T>0$, reaching $s'$ at $T$ first requires to start from an initial state $s$ (chosen according to $\iota$), followed by some delay $t$ (chosen according to $\mu(s)$), leaving the system in the state $F(t,s)$.  From this state it remains to reach $s'$ within time $(T-t)$ using $\eta(F(t,s))$ as initial state density.

\subsubsection{Hybrid Markov Decision Processes}
A hybrid Markov decision process (HMDP) is essentially an EMDP where the actions of the agent are selected according to some time period $P\in\realnumber_{\geq0}$, and  where the next-state probability density function $T$ is obtained from an SHS.

\begin{definition}[Hybrid Markov decision process]\label{def:HDMP}
A \emph{hybrid Markov decision process} is a tuple ${\cal HM} = (\statespace, \state_0, \act, P, N, {\cal H}, \costfunction, \goal)$ where $\statespace, \state_0, \act, \costfunction, \goal$ are defined the same way as for an EMDP, and
\begin{itemize}
    \item $P\in\realnumber_{\geq0}$ is the period of the agent,
    \item $N:\statespace\times\act\rightarrow (\statespace\rightarrow \realnumber_{\geq0})$ maps each state $s$ and action $a$ to a probability density function determining the immediate next state under $a$, and
    \item ${\cal H}=(\statespace,F,\mu,\eta)$ is a stochastic hybrid system describing the responses of the environment.
\end{itemize}
\end{definition}

An HMDP ${\cal HM} = (\statespace, \state_0, \act, P, N, {\cal H}, \costfunction, \goal)$ induces the EMDP
${\cal M}_{{\cal HM}}=$ $(\statespace, \state_0, \act, \transitionfunction, \costfunction, \goal)$, where $\transitionfunction$ is given by
$\transitionfunction(s,a) = \Delta_{{\cal H},N(s,a)}^P$.
That is, the next-state probability density function of ${\cal M}_{{\cal HM}}$ is given by the state density after a delay of $P$ (the period) according to ${\cal H}$ with initial state density $N$.

\begin{figure}[tb]
    \centering
    \begin{subfigure}[b]{0.49\textwidth}
        \centering
        \includesvg[width=\linewidth]{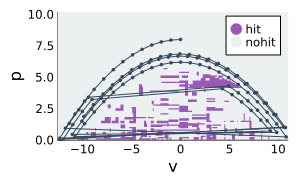}
        \caption{Strategy.}
    \end{subfigure}
    \hfill
    \begin{subfigure}[b]{0.49\textwidth}
        \centering
        \includesvg[width=\linewidth]{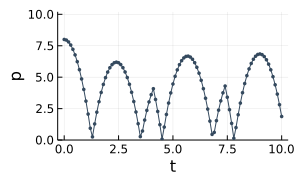}
        \caption{Example run for 10 seconds.}
    \end{subfigure}
    \caption{Near-optimal strategy learned by \uppaalstratego.}
    \label{fig:UnshieldedTrace}
\end{figure}

\begin{wrapfigure}{r}{0.60\textwidth}
        \centering
        \includegraphics[width=\linewidth]{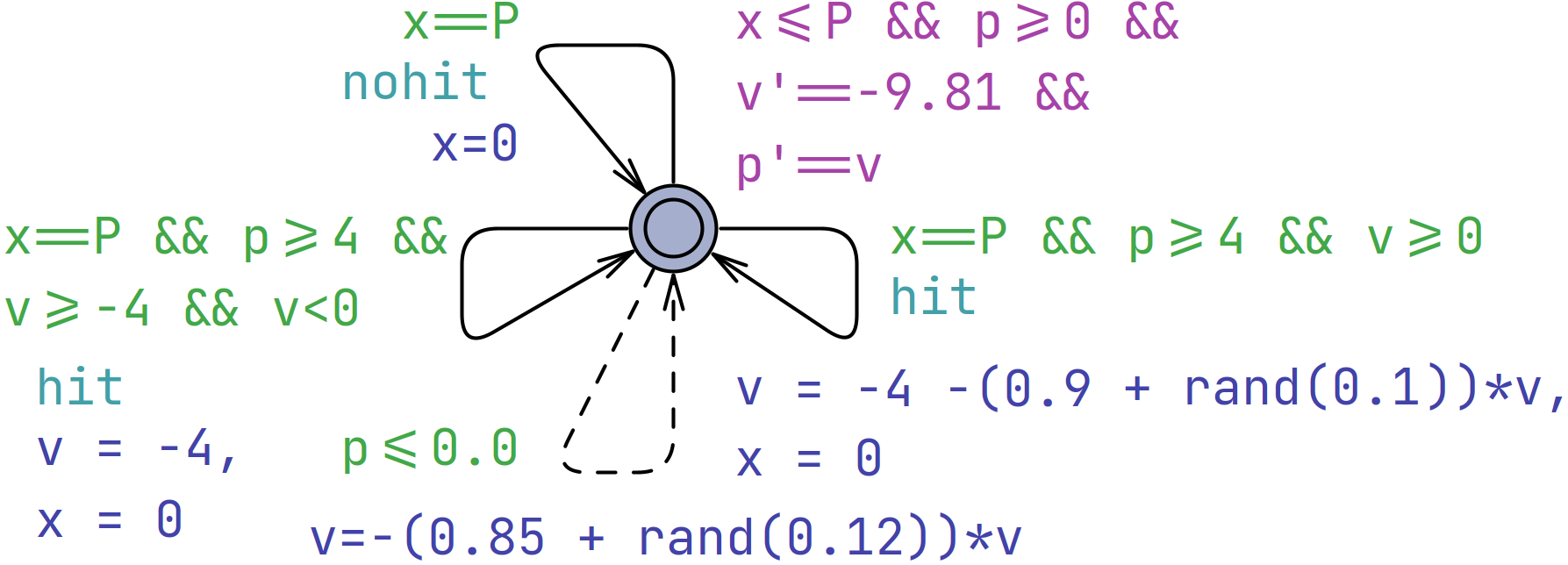}
        \caption{An HMDP for hitting a bouncing ball.}
        \vspace{-15pt}
        \label{fig:HBB}
    \end{wrapfigure}

\refstepcounter{examplectr}
\paragraph{Example~{\theexamplectr}\label{ex:hitting} (Hitting the bouncing ball).}
    Fig.~\ref{fig:HBB} shows an HMDP extending the SHS of the bouncing ball from  Fig.~\ref{fig:BBBehaviour}(a). Now a player has to keep the ball bouncing indefinitely by periodically choosing between the actions {\sl hit} and {\sl nohit},
    (three solid transitions).
    The period $P = 0.1$ is modeled by a clock $x$ with suitable invariant, guards and updates.
    The top transition triggered by the {\sl nohit} action has no effect on the state (but will have no cost).
    The {\sl hit} action affects the state only if the height of the ball is at least 4m ($p\geq 4$).
    The left transition applies if the ball is falling with a speed not greater than $-4$m/s ($v\geq-4$) and accelerates to a velocity of $-4$m/s.
    The right transition applies if the ball is rising, and sets the velocity to a random value in $[-v-4,-0.9v-4]$.
    The bottom dashed transition represents the bounce of the ball as in Fig.~\ref{fig:BBBehaviour}(a), which is part of the environment and outside the control of the agent.

    A time-extended state $(p,v,t)$ is in the goal set ${\cal G}$ if either $t\geq 120$ or $(p\leq 0.01 \wedge |v|\leq 1)$ (the ball is deemed dead).
    The cost ($C$) is 1 for the {\sl hit} action and 0 for the {\sl nohit} action, with an additional penalty of $1{,}000$ for transitions leading to a dead state.
    Fig.~\ref{fig:UnshieldedTrace} illustrates the near-optimal strategy $\sigma^*$ obtained by the RL method implemented in \uppaalstratego and the prefix of a random run.
    The expected number of {\sl hit} actions of $\sigma^*$ within 120s is approximately $48$.\exampleend

%% file: computingreachability.tex
\section{Safety, Partitioning, Synthesis and Shielding}\label{sec:computingreachability}
\subsubsection{Safety}
In this section we are concerned with a strategy obtained for a given EMDP being {\sl safe}.
For example, a safety strategy for hitting the bouncing ball must ensure that the ball never reaches a dead state ($p\leq 0.01 \wedge |v|\leq 1$).
In fact, although safety was encouraged by cost-tweaking, the strategy $\sigma^*$  in Fig.~\ref{fig:UnshieldedTrace} is {\sl not} safe.
In the following we use symbolic techniques to synthesize safety strategies.

Let $\curlym = (\statespace, \state_0, \act, \transitionfunction, \costfunction, \curlyg)$ be an EMDP.
A safety property $\varphi$ is a set of states $\varphi\subseteq\statespace$.
A run $\pi=s_0a_0s_1a_1s_2\dots$ is safe with respect to $\varphi$ if $s_i\in\varphi$ for all $i\geq 0$.
Given a nondeterministic strategy $\sigma:\statespace\rightarrow2^{\act}$, a run $\pi=s_0a_0s_1a_1s_2\dots$ of ${\cal M}$ is an outcome of $\sigma$ if $a_i\in \sigma(s_i)$  for all $i$.
We say that $\sigma$ is a safety strategy with respect to $\varphi$ if all runs that are outcomes of $\sigma$ are safe.

\subsubsection{Partitioning and Strategies}
Given the infinite-state nature of the EMDP ${\cal M}$, we will resort to finite partitioning (similar to \cite{ZikelicLHC23}) of the state space in order to algorithmically synthesize nondeterministic safety strategies.
Given a predefined granularity $\granularity$, we partition the state space into disjoint regions of equal size (we do this for simplicity; our method is independent of the particular choice of the partitioning).
The partitioning along each dimension of $\statespace$ is a half-open interval
belonging to the set $\Idelta = \lbrace \rinterval{k \granularity}{k \granularity + \granularity} \mid k \in \integers \rbrace$.
For a bounded $k$-dimensional state space $\statespace$, $\partitioning = \lbrace \partition \in \Idelta^k \mid \partition \cap \statespace \neq \emptyset \rbrace$ provides a finite partitioning of $\statespace$ with granularity $\granularity$. For each $s\in\statespace$ we denote by $[s]_{{\cal A}}$ the unique region containing $s$.

Given an EMDP ${\cal M}$, a partitioning ${\cal A}$ induces a finite labeled transition system $\transitionsystem=({\partitioning},\act,\rightarrow)$, where
\[\mu\xrightarrow[]{a}\mu' \iff \exists s\in\mu.\ \exists s'\in\mu'.\ T(s,a)(s')>0.\]

\begin{figure}[tb]
    \centering
    \begin{subfigure}[b]{0.49\textwidth}
        \centering
        \includesvg[width=\linewidth]{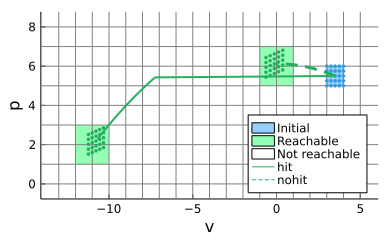}
        \caption{Scenario where the ball is rising and high enough to be hit.}
    \end{subfigure}
    \hfill
    \begin{subfigure}[b]{0.49\textwidth}
        \centering
        \includesvg[width=\linewidth]{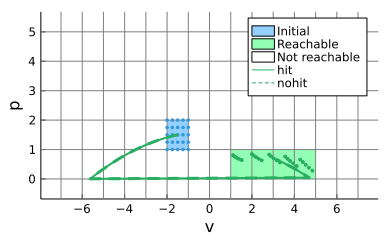}
        \caption{Scenario where the ball is too low to be hit, but bounces off the ground.}
    \end{subfigure}
    \caption{State-space partitioning for Example~\ref{ex:hitting}.
    Starting in the blue region and depending on the action, the system can end up in the green regions within one time period, witnessed by simulations from $16$ initial states.
    }
    \label{fig:SquaresReachabilityBarbaricGroup}
    \label{fig:SquaresReachabilityGroup}
\end{figure}

Fig.~\ref{fig:SquaresReachabilityGroup} shows a partitioning for the running example and displays some witnesses for transitions in the induced transition system.

Next, we view $\transitionsystem$ as a 2-player game.
For a region $\mu\in{\cal A}$, Player~1 challenges with an action $a\in\act$.
Player~2 responds with a region $\mu'\in{\cal A}$ such that $\mu\xrightarrow{a}\mu'$.

\begin{definition}[Safe regions]
Let $\varphi\subseteq \statespace$ be a safety property and ${\cal A}$ a partitioning. 
We denote by $\varphi^{\cal A}$ the set $\{\mu\in{\cal A} \mid \mu\subseteq\varphi\}$. 
The set of safe regions with respect to $\varphi$ is the maximal set of regions $\mathbb{S}_\varphi$ such that
\begin{equation}
\label{defeq}
\mathbb{S}_\varphi = \varphi^{\cal A} \cap \{ \mu \mid \exists a.\ \forall \mu'.\ \mu\xrightarrow{a}\mu' \implies \mu'\in \mathbb{S}_\varphi \}.
\end{equation}
\end{definition}

Given the finiteness of ${\cal A}$ and monotonicity of~\eqref{defeq}, $\mathbb{S}_\varphi$ may be obtained in a finite number of interations using Tarski's fixed-point theorem~\cite{Tarski55}.

A (nondeterministic) strategy for $\transitionsystem$ is a function $\nu:{\cal A}\rightarrow 2^{\act}$.
The most permissive safety strategy $\nu_\varphi$ obtained from $\mathbb{S}_\varphi$~\cite{BernetJW02} is given by
\[\nu_\varphi(\mu)=\{a\mid \forall\mu'.\ \mu\xrightarrow{a}\mu'\implies \mu'\in\mathbb{S}_\varphi\}.\]

The following theorem states that we can obtain a safety strategy for the original EMDP ${\cal M}$ from a safety strategy $\nu$ for $\transitionsystem$.

\begin{theorem}\label{thm:safety_transfer}
Given an EMDP ${\cal M}$, safety property $\varphi\subseteq \statespace$ and partitioning ${\cal A}$, if $\nu$ is a safety strategy for $\transitionsystem$, then $\sigma(s)=\nu([s]_\partitioning)$ is a safety strategy for ${\cal M}$.
\end{theorem}

\subsubsection{Approximating the 2-player Game}

Let ${\cal M}$ be an EMDP  and $\varphi$ be a safety property.
To algorithmically compute the set of safe regions $\mathbb{S}_\varphi$ for a given partitioning ${\cal A}$, and subsequently the most permissive safety strategy $\nu_\varphi$, the transition relation $\xrightarrow{a}$ needs to be a decidable predicate.
If ${\cal M}$ is derived from an HMDP ${\cal HM} = (\statespace, \state_0, \act, P, N, {\cal H}, \costfunction, \curlyg)$, this requires decidability of the predicate $\Delta_{{\cal H},N(s,a)}^P(s')>0$.
Consider the bouncing ball from Example~\ref{ex:hitting}.
The regions are of the form $\mu=\{(p,v)\mid l_p\leq p <u_p \wedge l_v\leq v<u_v\}$.
For given regions $\mu, \mu'$, the predicate $\mu\xrightarrow{nohit}\mu'$ is equivalent to the following first-order predicate over the reals (note that $F((p,v),t)$ is a pair of polynomials in $p, v$ and $t$):\footnote{We assume that at most one bounce can take place within the period $P$.}
\begin{align*}
    \exists (p,v)\in\mu.\ F((p,v),P)\in\mu' \vee {} & \exists \beta\in[0.85,0.97].\ \exists t'\leq P.\ \exists v'. \\
            &F((p,v),t')=(0,v') \wedge F((0,-\beta\cdot v'),P-t')\in\mu'
\end{align*}

For this simple example, the validity of the formula can be decided~\cite{Tarski48}, which may however require doubly exponential time~\cite{DavenportH88}, and worse, when considering nonlinear dynamics with, e.g., trigonometric functions, the problem becomes undecidable~\cite{Laczkovich03}.
One can obtain a conservative answer via over-approximate reachability analysis~\cite{DFPP18}; in Section~\ref{sec:experiments} we compare to such an approach and demonstrate that, while effective, that approach also does not scale.
This motivates to use an efficient and robust alternative.
We propose to approximate the transition relation using equally spaced samples, which are simulated according to the SHS ${\cal H}$ underlying the given HMDP ${\cal HM}$.

\begin{wrapfigure}{r}{0.5\textwidth}
    \vspace{-45pt}
    \begin{minipage}{\linewidth}
        \begin{algorithm}[H]
        \begin{algorithmic}[1]
    \Input{$\mu \in {\cal A}, a \in \act$}
    \Output{$\mu\xrightarrow{a}_{app}\mu'$ iff $\mu' \in R$}
    \State $R = \emptyset$
    \ForAll{$s_i \in  app[\mu]$}
          \State select $s_i' \sim N(s_i,a)$
          \State simulate ${\cal H}$ from $s_i'$ for $P$ time units
          \State let ${s_i''}$ be the resulting state
          \State add $[s_i'']_{\cal A}$ to $R$
      \EndFor
        \end{algorithmic}
        \caption{Approximation of $\xrightarrow{a}$}
        \label{MCAlgo}
        \end{algorithm}
    \end{minipage}
    \vspace{-20pt}
\end{wrapfigure}
Algorithm~\ref{MCAlgo} describes how to compute an approximation $\mu\xrightarrow{a}_{app}\mu'$ of $\mu\xrightarrow{a}\mu'$.
The algorithm draws from a finite set of $n$ evenly distributed supporting points per dimension $app[\mu]=\{s_1,\dots,s_{n^k}\} \subseteq \mu$ and simulates ${\cal H}$ for $P$ time units.
A region $\mu'$ is declared reachable from $\mu$ under action $a$ if it is reached in at least one simulation.
When stochasticity is involved in a simulation, additional care must be taken. The random variables can be considered an additional dimension to be sampled from; alternatively, a worst-case value can be used if available, such as the bouncing ball with the highest velocity damping.
Fig.~\ref{fig:SquaresReachabilityBarbaricGroup} illustrates 16 ($n=4$) possible starting points for the bouncing ball together with most pessimistic outcomes, depending on the action taken.

The result $\xrightarrow{a}_{app}$ is an underapproximation of the transition relation $\xrightarrow{a}$
, with a corresponding transition system $\approxts = (\partitioning,\act,\xrightarrow{}_{app})$. 
Thus if we compute a safety strategy $\nu$ from $\xrightarrow{a}_{app}$, then the strategy $\sigma(s)=\nu([s]_\partitioning)$ from Theorem~\ref{thm:safety_transfer} is not necessarily safe.
However, in Section~\ref{sec:experiments} we will see that this strategy is statistically safe in practice.
We attribute this to two reasons.
1)~The underapproximation of $\xrightarrow{a}_{app}$ can be made accurate.
2)~Since $\xrightarrow{a}$ is defined over an abstraction, it is often robust against small approximation errors.

%% file: shield.tex
\subsubsection{Shielding}
As argued above, we can obtain the most permissive safety strategy $\nu_\varphi$ from $\xrightarrow{a}_{app}$ over ${\partitioning}$ and then use $\sigma_\varphi(s)=\nu_\varphi([s]_\partitioning)$ as an approximation of the most permissive safety strategy over the original HMDP.
We can employ $\sigma_\varphi$ to build a shield.
As discussed in the introduction, we focus on two ways of shielding: \textit{pre-shielding} and \textit{post-shielding} (recall Fig.~\ref{fig:shieldingTypes}).
In pre-shielding, the shield is already active during the learning phase of the agent, which hence only trains on sets of safe actions.
In post-shielding, the shield is only applied after the learning phase, and unsafe actions chosen by the agent are corrected (which is possibly detrimental to the performance of the agent).

\begin{figure}[tb]
    \centering
    \begin{minipage}{0.49\linewidth}
        \centering
        \includesvg[width=\linewidth]{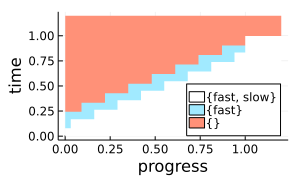}
    \end{minipage}
    \hfill
    \begin{minipage}{0.49\linewidth}
        \centering
        \includesvg[width=\linewidth]{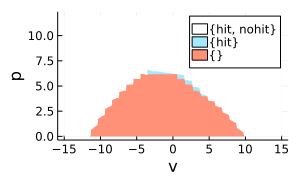}
    \end{minipage}
    \caption{Synthesized strategies for random walk (left) and bouncing ball (right).}
    \label{fig:ShieldsRWBB}
\end{figure}

Fig.~\ref{fig:ShieldsRWBB} shows examples of such strategies for the random walk (Example~\ref{ex:random_walk}) and the bouncing ball.
As can be seen, most regions of the state space are either unsafe (black) or both actions are safe (white). Only in a small area (purple) will the strategy enforce walking fast or hitting the ball, respectively.
In the white area, the agent can learn the action that leads to the highest performance.

\begin{figure}[tb]
    \centering
    \includesvg[width=\linewidth]{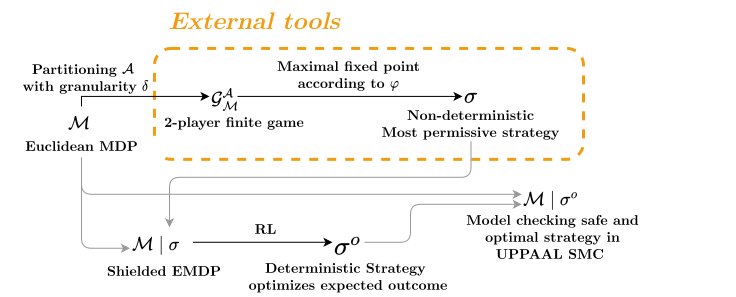}
    \caption{Complete method for pre-shielding and statistical model checking (SMC).}
    \label{fig:StrategoMyPreShielding}
\end{figure}

We use \uppaalstratego~\cite{stratego} to train a shielded agent based on $\sigma_\varphi$.
The complete workflow of pre-shielding and learning is depicted in Fig.~\ref{fig:StrategoMyPreShielding}.
Starting from the EMDP, we partition the state space, obtain the transition system using Algorithm~\ref{MCAlgo} and solve the game according to a safety property $\varphi$, as described above.
The produced strategy is then conjoined with the original EMDP to form the shielded EMDP, and reinforcement learning is used to produce a near-optimal deterministic strategy $\sigma^*$.
This strategy can then be used in the real world, or get evaluated via statistical model checking.
The only difference in the workflow in post-shielding is that the strategy $\sigma_\varphi$ is not applied to the EMDP, but on top of the deterministic strategy $\sigma^*$.

%% file: results.tex

\section{Experiments}
\label{sec:experiments}

In this section we study our proposed approach with regard to different aspects of our shields.
In addition to the random walk (Example~\ref{ex:random_walk}) and bouncing ball (Example~\ref{ex:hitting}), we consider three benchmark cases:
\begin{itemize}
    \item \emph{Cruise control}~\cite{larsen2015cruisecontrol,10.1007/978-3-030-23703-5_6,DBLP:conf/qest/AshokKLCTW19}:
    A car is controlled to follow another car as closely as possible without crashing.
    Either car can accelerate, keep its speed, or decelerate freely, which makes finding a strategy challenging.
    This model was subject to several previous studies where a safety strategy was carefully designed, while our method can be directly applied without human effort.

    \item \emph{DC-DC converter}~\cite{dcdcconverter}:
    This industrial DC-DC boost converter transforms input voltage of $10$V to output voltage of $15$V.
    The controller switches between storing energy in an inductor and releasing it.
    The output must stay in $\pm0.5$V around $15$V, and the amount of switching should be minimized.

    \item \emph{Oil pump}~\cite{hydac}:
    In this industrial case, flow of oil into an accumulator is controlled to satisfy minimum and maximum volume constraints, given a consumption pattern that is piecewise-constant and repeats every $20$~seconds.
    Since the exact consumption is unknown, a random perturbation is added to the reference value. To reduce wear, the volume should be kept low. 
\end{itemize}

\begin{figure}[tb]
    \centering
    \begin{minipage}{.49\textwidth}
        \begin{subfigure}{\linewidth}
            \centering
            \includesvg[width=\linewidth]{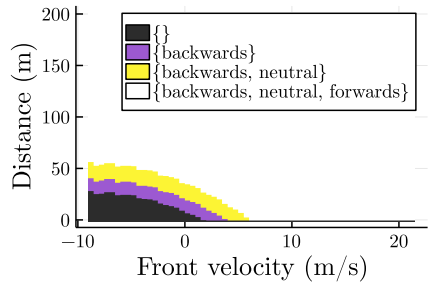}
            \caption{Cruise control ($n=4$, $\granularity=0.5$) when the car's velocity is $0m/s$}
            \label{fig:CCShield}
        \end{subfigure}
        \\
        \begin{subfigure}{\linewidth}
            \centering
            \includesvg[width=\linewidth]{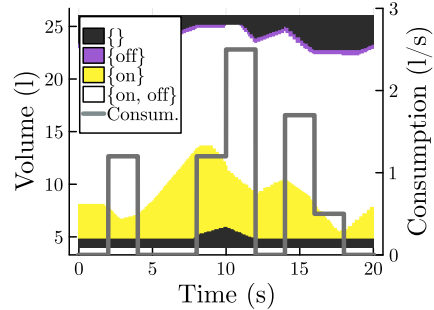}
            \caption{DC-DC boost converter ($n=4$, $\granularity=0.01$) when the output resistance is $30\Omega$.}
            \label{fig:DCShield}
        \end{subfigure}
    \end{minipage}
    \hfill
    \begin{minipage}{.49\textwidth}
        \begin{subfigure}{\linewidth}
            \centering
            \includesvg[width=\linewidth]{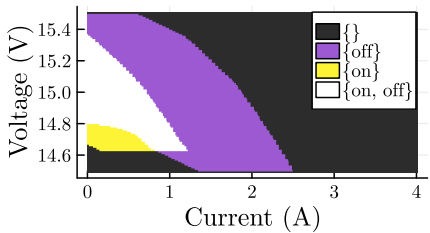}
            \caption{Oil pump ($n=4, \granularity=0.1$) when the pump is \textit{on}. The periodic piecewise consumption pattern has been overlaid.
            Turning off the pump requires it to stay off for two seconds, which could cause an underflow in the yellow area.
            Conversely, the purple area shows the states where the pump \textit{must} be turned off to avoid overflow.
            Since the pump is on in this projection, this can wait until the last moment.}
            \label{fig:OPShieldOn}
        \end{subfigure}
    \end{minipage}
    \caption{Projected views of synthesized most permissive safety strategies.}
    \label{fig:Shields}
\end{figure}

Fig.~\ref{fig:Shields} shows the synthesized most permissive safety strategies.
For instance, in Fig.~\ref{fig:CCShield} we see the strategy for the cruise-control example when the controlled car is standing still. If the car in front is either close or reverses at high speed, the controlled car must also reverse (purple area).
The yellow area shows states where it is safe to stand still but accelerating may lead to a collision.

We conduct four series of experiments to study different aspects of our approach.
\begin{enumerate*}[label={(\arabic*)}]
    \item The quality of our approximation of the transition relation $\xrightarrow{a}_{app}$,\label{lab:exp:bbtrans}
    \item the computational performance of our approximation in comparison with a fully symbolic approach,\label{lab:exp:julia}
    \item the performance in terms of reward and safety of the pre- and post-shields synthesized with our method, and\label{lab:exp:prepost}
    \item the potential of post-optimization for post-shielding.\label{lab:exp:postopt}
\end{enumerate*}

All experiments are conducted on an AMD Ryzen 7 5700x with 33~GiB RAM.
Our implementation is written in Julia, and we use \uppaalstratego~\cite{stratego} for learning and statistical model checking.
The experiments are available online~\cite{REP}.

\subsubsection{Quality of the Approximated Transition System}
In the first experiment we statistically assess the approximation quality of $\xrightarrow{a}_{app}$ wrt.\ the underlying infinite transition system.
For varying granularity $\granularity$ of $\partitioning$ and numbers of supporting points $n$ per dimension (see Section~\ref{sec:computingreachability}) we first compute $\xrightarrow{a}_{app}$ with Algorithm~\ref{MCAlgo}.
Then we uniformly sample $10^8$ states $\state$ and compute their successor states $\state'$ under a random action $a$.
Finally we count how often $[\state]_\partitioning \xrightarrow{a}_{app} [\state']_\partitioning$ holds.

\begin{figure}[t]
    \centering
    \begin{minipage}{0.49\linewidth}
        \centering
        \includesvg[width=1\linewidth]{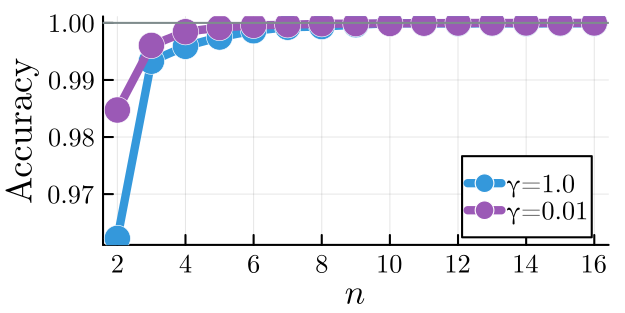}
    \end{minipage}
    \hfill
    \begin{minipage}{0.49\linewidth}
        \centering
        \includesvg[width=1\linewidth]{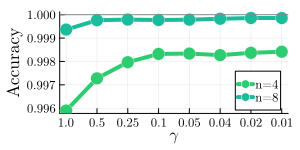}
    \end{minipage}
    \caption{Accuracy of the approximation $\xrightarrow{\action}_{app}$ under different granularity $\granularity$ and number of supporting points $n$ per dimension.}
    \label{fig:BarbaricAccuracy}
\end{figure}

Here we consider the bouncing-ball model, where we limit the domain to $p\in[0,15]$, $v\in[-15,15]$.
The results are shown in Fig.~\ref{fig:BarbaricAccuracy}.
An increase in the number of supporting points $n$ correlates with increased accuracy.
For $\granularity \leq 1$, using $n=3$ supporting points already yields accuracy above $99$\%.
Finer partition granularity $\granularity$ increases accuracy, but less so compared to increasing $n$.

\subsubsection{Comparison with Fully Symbolic Approach}

As described in Section~\ref{sec:computingreachability}, as an alternative to Algorithm~\ref{MCAlgo} one can use a reachability algorithm to obtain an overapproximation of the transition relation $\xrightarrow{a}$.
Here we analyze the performance of such an approach based on the reachability tool \juliareach~\cite{JuliaReach}.
Given a set of initial states of a hybrid automaton where we have substituted probabilities by nondeterminism, \juliareach can compute an overapproximation of the successor states.
In \juliareach, we select the reachability algorithm from~\cite{GuernicG09}. This algorithm uses time discretization, which requires a small time step to give precise answers. This makes the approach expensive. For instance, for the bouncing-ball system, the time period is $P = 0.1$ time units, and a time step of $0.001$ time units is required, which corresponds to $100$ iterations.

The shield obtained with \juliareach is safe by construction.
To assess the safety of the shield obtained with Algorithm~\ref{MCAlgo}, we choose an agent that selects an action at random and let it act under the post-shield for $10^6$ episodes.
(We use a random agent because a learned agent may have learned to act safely most of the time and thus not challenge the shield as much.)
If no safety violation was detected, we compute 99\% confidence intervals for the statistical safety.

\begin{table}[tb]
    \caption{
    Synthesis results for the bouncing ball under varying granularity ($\granularity$) and supporting points ($n$) using Algorithm~\ref{MCAlgo} (top) and two choices of the time-step parameter using \juliareach (bottom).
    The safety probability is computed for a 99\% confidence interval.
    $\granularity=0.02$ corresponds to $9.0\cdot10^5$ partitions, and $\granularity=0.01$ quadruples the number of partitions to $3.6\cdot10^6$.
    }
    \label{tab:BBSynthesis}
    \centering
    \include{graphics/BBSynthesis.tex}
\end{table}

We consider again the bouncing-ball model.
\juliareach requires a low partition granularity $\granularity = 0.01$; for $\granularity = 0.02$ it cannot prove that a safety strategy exists, which may be due to conservatism, while our method is able to synthesize a shield that, for $n \geq 4$, is statistically safe.
Table~\ref{tab:BBSynthesis} shows the results obtained from the two approaches.
In addition, the reachability algorithm uses time discretization, and a small time step is required to find a safety strategy.

\begin{wrapfigure}{r}{0.41\textwidth}
    \centering
    \includesvg[width=\linewidth]{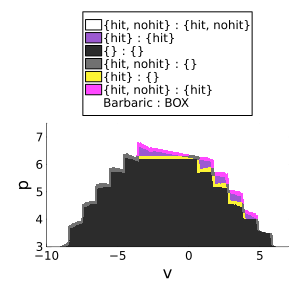}
    \caption{Superimposed strategies of our method and \juliareach.}
    \vspace{-15pt}
    \label{fig:DifferenceRigorousBarbaric}
\end{wrapfigure}
We remark that the bouncing-ball model has linear dynamics, for which reachability analysis is relatively efficient compared to nonlinear dynamics, and thus this model works in favor of the symbolic method.
However, the hybrid nature of the model and the large number of queries (one for each partition-action pair) still make the symbolic approach expensive.
Considering the case $\granularity = 0.01$ and $n = 4$, our method can synthesize a strategy in $19$~minutes, while the approach based on \juliareach takes $41$~hours.
%

Fig.~\ref{fig:DifferenceRigorousBarbaric} visualizes the two strategies and shows how the two approaches largely agree on the synthesized shield -- but also the slightly more pessimistic nature of the transition relation computed with \juliareach.

\subsubsection{Evaluation of Pre- and Post-shields}

In the next series of experiments, we evaluate the full method of obtaining a shielded agent.
The first step is to approximate $\xrightarrow{a}_{app}$ using Algorithm~\ref{MCAlgo} and extract the most permissive safety strategy $\sigma_\varphi$ to be used as a shield.
For the second step we have two options: pre- or post-shielding.
Recall from Fig.~\ref{fig:shieldingTypes} that a pre-shield is applied to the agent during training while a post-shield is applied after training.

\begin{figure}[t!]
    \centering
    \begin{subfigure}[b]{0.49\textwidth}
        \centering
        \includesvg[width=\linewidth]{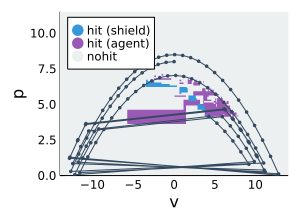}
        \caption{Pre-shield.}
    \end{subfigure}
    \hfill
    \begin{subfigure}[b]{0.49\textwidth}
        \centering
        \includesvg[width=\linewidth]{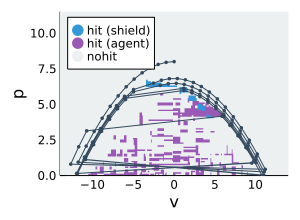}
        \caption{Post-shield.}
    \end{subfigure}
     \caption{Learned shielded strategies for the bouncing ball.
     }
    \label{fig:ShieldedTrace}
\end{figure}

In the case of the bouncing ball, the post-shielded agent's strategy is shown in Fig.~\ref{fig:ShieldedTrace}(b).
It consists of the unshielded strategy from Fig.~\ref{fig:UnshieldedTrace} plus the purple regions of the safety strategy in Fig.~\ref{fig:ShieldsRWBB}(b).
Correspondingly, Fig.~\ref{fig:ShieldedTrace}(a) shows the pre-shielded strategy, which is significantly simpler because it does not explore unsafe regions of the state space.
This also leads to faster convergence.

Table~\ref{tab:synthesis} reports the same data as in Table~\ref{tab:BBSynthesis} for the other models.
Overall, we see a similar trend in all tables.
For a low number of supporting points (say, $n = 3$) we can obtain a safety strategy that we find to be statistically safe.
In all cases, no unsafe run was detected in the statistical evaluation.
The synthesis time varies depending on the model and is generally feasible.
The longest computation times are seen for the oil-pump example, which has the most dimensions.
Still, times are well below \juliareach for the comparatively simple bouncing ball.

\begin{table}[tb!]
    \caption{
        Shield synthesis for different models and granularities $\granularity$ computed using Algorithm~\ref{MCAlgo}.
        The safety probability is computed for a 99\% confidence interval.
    }
    \label{tab:synthesis}
    \begin{minipage}[t]{.49\textwidth}
        \begin{subtable}[t]{\textwidth}
            \caption{Cruise control. $\granularity=1$ corresponds to $1.9\cdot10^5$ partitions, and $\granularity=0.5$ to $1.5\cdot10^6$.}
            \label{tab:CCSynthesis}
            \centering
            \include{graphics/CCSynthesis.tex}
        \end{subtable}
        \\[-\baselineskip]
        \stepcounter{subtable}
        \begin{subtable}{\textwidth}
            \caption{Oil pump. $\granularity=0.2$ corresponds to $2.8\cdot10^5$ partitions, and $\granularity=0.1$ to $1.1\cdot10^6$.}
            \label{tab:OPSynthesis}
            \centering
            \include{graphics/OPSynthesis.tex}
        \end{subtable}
    \end{minipage}
    \hfill
    \begin{minipage}[t]{.49\textwidth}
        \addtocounter{subtable}{-2}
        \begin{subtable}[t]{\textwidth}
            \caption{DC-DC boost converter. $\granularity=0.05$ corresponds to $3.1\cdot10^5$ partitions, $\granularity=0.02$ to $1.7\cdot10^6$ and $\granularity=0.01$ to $7.0\cdot10^6.$}
            \label{tab:DCSynthesis}
            \centering
            \include{graphics/DCSynthesis.tex}
        \end{subtable}
        \addtocounter{subtable}{2}
    \end{minipage}
\end{table}

\triplesvgspace{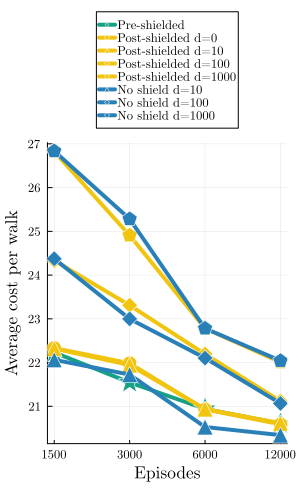}{Average cost per run.}
{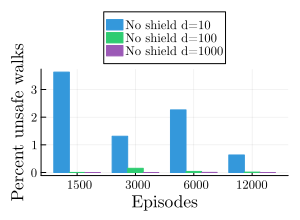}{Safety violations for unshielded agents}
{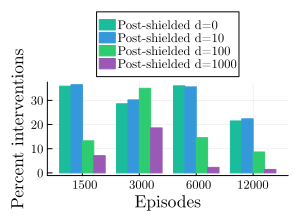}{Interventions for post-shielded agents.}
{RWShieldingResultsGroup}{Results of shielding the random walk using $\granularity=0.005$.}{tb!}

\triplesvgspace{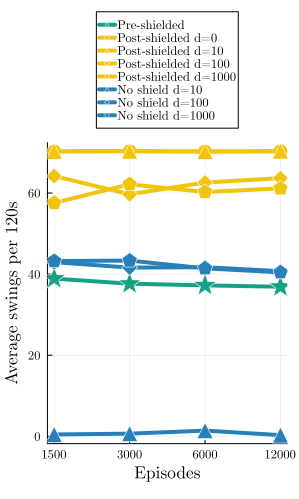}{Average \emph{hit} actions per 120s.}
{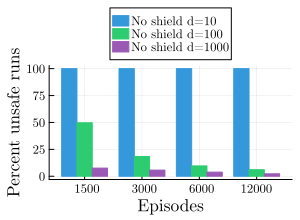}{Safety violations for unshielded agents.}
{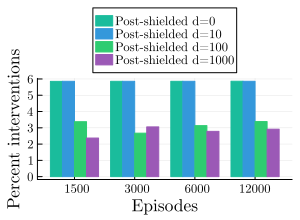}{Interventions for post-shielded agents.}
{BBShieldingResultsGroup}{Results of shielding the bouncing ball using $n=16$, $\granularity=0.01$.}{tb!}

\triplesvgspace{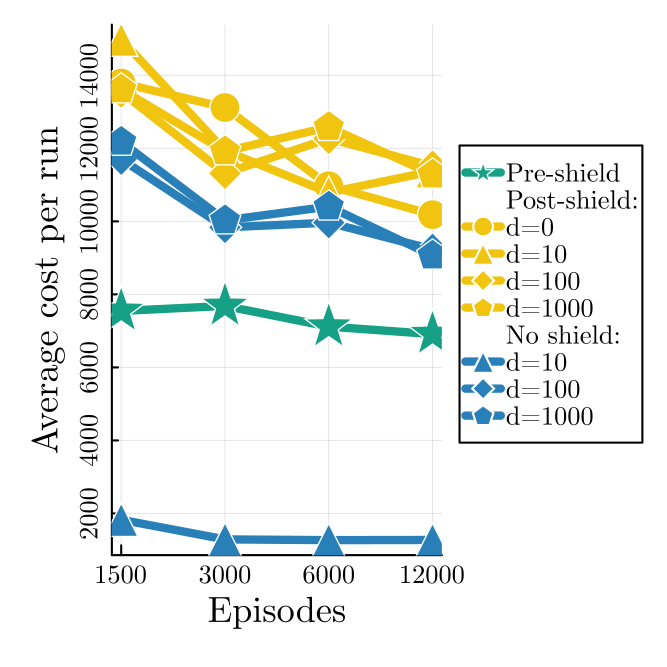}{Accumulated distance per 120s.}
{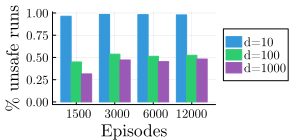}{Safety violations for unshielded agents.}
{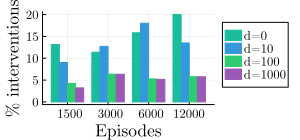}{Interventions for post-shielded agents.}
{CCShieldingResultsGroup}{Results of shielding the cruise control using $n=4$, $\granularity=0.5$.}{tb!}

\triplesvgspace{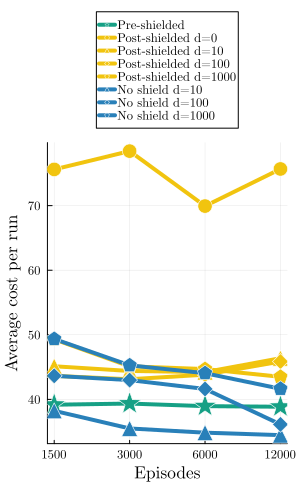}{Accumulated error plus number of switches per 120{\textmu}s.}
{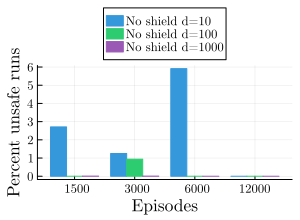}{Safety violations for unshielded agents.}
{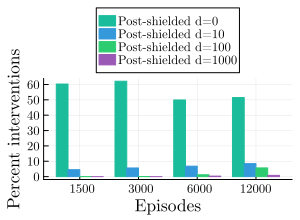}{Interventions for post-shielded agents.}
{DCShieldingResultsGroup}{Results of shielding the DC-DC boost converter using $n=4$, $\granularity=0.01$.}{tb!}

\triplesvgspace{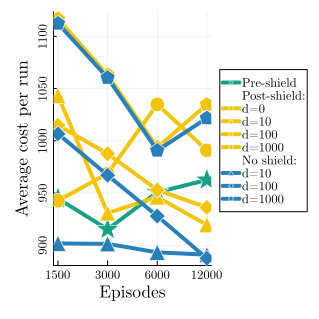}{Accumulated oil volume per 120s.}
{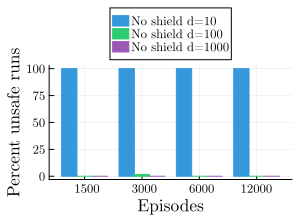}{Safety violations for unshielded agents.}
{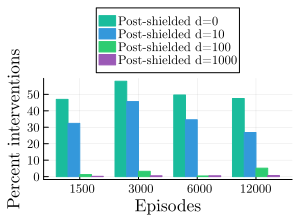}{Interventions for post-shielded agents.}
{OPShieldingResultsGroup}{Results of shielding the oil pump using $n=4$, $\granularity=0.1$.}{tb!}

\smallskip

Next, we compare our method to other options to make an agent safe(r).
As the baseline, we use the classic RL approach, where safety is encouraged using reward shaping.
We experiment with a deterrence $d \in \{0, 10, 100, 1000\}$ (negative reward) as a penalty for safety violations for the learning agent.
Note that this penalty is only applied during training, and not included in the total cost when we evaluate the agent below.
As the second option, we use a post-shielded agent, to which the deterrence also applies.
The third option is a pre-shielded agent.
In all cases, training and evaluation is repeated 10 times, and the mean value is reported.
The evaluation is based on 1000 traces for each repetition.

\vspace{20pt}

Figures~\ref{fig:RWShieldingResultsGroup}, to~\ref{fig:OPShieldingResultsGroup}  report the results for the different models.
Each subfigure shows the following content:
(a)~shows the average cost of the final agent, (b)~shows the amount of safety violations of the unshielded agents and (c)~shows the number of times the post-shielded agents were intervened by the shield.

Overall, we observe similar tendencies. The unshielded agent has lowest average cost at deployment time under low deterrence, but it also violates safety.
Higher deterrence values improve safety, but do not guarantee it.

The pre-shielded agents outperform the post-shielded agents.
This is because they learn a near-optimal strategy subject to the shield, while the post-shielded agents may be based on a learned unsafe strategy that contradicts the shield, and thus the shield interference can be more detrimental.

\begin{table}[tb]
    \caption{Change of post-optimization relative to the uniform-choice strategy.
        The strategy was trained for $12{,}000$ episodes with $d=10$ and post-optimized for $4{,}000$ episodes.
        Performance of the pre-shielded agent is included for comparison, but interventions are not applicable (because the shield was in place during training).}
    \label{tab:CCPostShieldComparison}
    \centering
    \include{graphics/CCPostShieldComparison.tex}
\end{table}

\subsubsection{Post-Shielding Optimization}
When a post-shield intervenes, more than one action may be valid.
This leaves room for further optimization, for which we can use \uppaalstratego.
Compared to a uniform baseline, we assess three ways to resolve nondeterminism:
\begin{enumerate*}[label={\arabic*)}]
	\item minimizing interventions,
	\item minimizing cost and
	\item at the preference of the shielded agent (the so-called Q-value~\cite{Watkins89}).
\end{enumerate*}

Table~\ref{tab:CCPostShieldComparison} shows the effect of post-optimization on the cost and the number of interventions for the cruise-control example.
Notably, cost is only marginally affected, but the number of shield interventions can get significantly higher.
The pre-shielded agent has lower cost than all post-optimized alternatives.

%% file: graphics/BBSynthesis.tex
\begin{tabular}{@{\hspace*{1mm}} c @{\hspace*{3mm}} c @{\hspace*{3mm}} c @{\hspace*{3mm}} r @{\hspace*{3mm}} c @{\hspace*{1mm}}}
\toprule
\textbf{$\granularity$}	&	\textbf{$\xrightarrow{a}_{app}$ method}	&	\textbf{Parameters}	&	\textbf{Time}	&	\textbf{Probability safe} \\
\midrule
0.02	&	\multirow{4}{*}{Algorithm~\ref{MCAlgo}}	&	$n=2$	&	1m\,50s	& \text{unsafe run found} \\ 
0.02	&		&	$n=4$	&	2m\,14s	&	$\interval{99.9999\%}{100\%}$	\\ 
0.02	&		&	$n=8$	&	4m\,02s	&	$\interval{99.9999\%}{100\%}$	\\ 
0.02	&		&	$n=16$	&	11m\,03s	&	$\interval{99.9999\%}{100\%}$	\\ 
\midrule
0.01	&	\multirow{4}{*}{Algorithm~\ref{MCAlgo}}	&	$n=2$	&	16m\,49s	&	$\interval{99.9999\%}{100\%}$	\\ 
0.01	&		&	$n=4$	&	19m\,00s 	&	$\interval{99.9999\%}{100\%}$	\\ 
0.01	&		&	$n=8$	&	27m\,21s	&	$\interval{99.9999\%}{100\%}$	\\ 
0.01	&		&	$n=16$	&	56m\,32s	&	$\interval{99.9999\%}{100\%}$	\\ 
\midrule
0.01	&	\multirow{2}{*}{\juliareach}	&	time step $0.002$	&	24h\,30m	&	considers $s_0$ unsafe	\\ 
0.01	&		&	time step $0.001$	&	41h\,05m	&	safe by construction	\\
\bottomrule
\end{tabular}

%% file: graphics/CCSynthesis.tex
\begin{tabular}{@{\hspace{0mm}} c @{\hspace{3mm}} c @{\hspace{3mm}} r @{\hspace{3mm}} c }
\toprule
\textbf{$\granularity$}	&	\textbf{$n$}	&	\textbf{Time}	&	\textbf{Probability safe}\\
\midrule
1	&	2	&	1m\,50s	&	Considers $s_0$ unsafe	\\
\midrule
0.5	&	2	&	13m\,16s	&	$\interval{99.9995\%}{100\%}$	\\
0.5	&	3	&	23m\,03s	&	$\interval{99.9995\%}{100\%}$	\\
0.5	&	4	&	35m\,55s	&	$\interval{99.9995\%}{100\%}$	\\
\bottomrule
\end{tabular}

%% file: graphics/OPSynthesis.tex
\begin{tabular}{@{\hspace{0mm}} c @{\hspace{3mm}} c @{\hspace{3mm}} r @{\hspace{3mm}} c }
    \toprule
    \textbf{$\granularity$}	&	\textbf{$n$}	&	\textbf{Time}	&	\textbf{Probability safe}\\
    \midrule
    0.2	&	2	&	3m\,07s	&	considers $s_0$ unsafe	\\
    \midrule
    0.1	&	2	&	32m\,15s	&	$\interval{99.9995\%}{100\%}$	\\
    0.1	&	3	&	1h\,37m	&	$\interval{99.9995\%}{100\%}$	\\
    0.1	&	4	&	5h\,23m	&	$\interval{99.9995\%}{100\%}$	\\
    \bottomrule
    \end{tabular}

%% file: graphics/DCSynthesis.tex
\begin{tabular}{@{\hspace{0mm}} c @{\hspace{3mm}} c @{\hspace{3mm}} r @{\hspace{3mm}} c }
    \toprule
    \textbf{$\granularity$}	&	\textbf{$n$}	&	\textbf{Time}	&	\textbf{Probability safe}\\
    \midrule
    0.05	&	2	&	41s	&	$\interval{99.9995\%}{100\%}$	\\
    0.05	&	3	&	1m\,50s	&	considers $s_0$ unsafe	\\
    0.05	&	4	&	3m\,30s	&	considers $s_0$ unsafe	\\
    \midrule
    0.02	&	2	&	3m\,43s	&	$\interval{99.9995\%}{100\%}$	\\
    0.02	&	3	&	8m\,59s	&	$\interval{99.9995\%}{100\%}$	\\
    0.02	&	4	&	18m\,11s	&	$\interval{99.9995\%}{100\%}$	\\
    \midrule
    0.01	&	2	&	15m\,48s	&	$\interval{99.9995\%}{100\%}$	\\
    0.01	&	3	&	38m\,26s	&	$\interval{99.9995\%}{100\%}$	\\
    0.01	&	4	&	1h\,19m	&	$\interval{99.9995\%}{100\%}$	\\
    \bottomrule
    \end{tabular}

%% file: graphics/CCPostShieldComparison.tex
\begin{tabular}{l @{\hspace*{5mm}} c c @{\hspace*{5mm}} c c} 
\toprule
\textbf{Configuration}	&	\multicolumn{2}{c}{\textbf{Cost}}\hspace*{1cm}	&	\multicolumn{2}{c}{\textbf{Interventions}}	\\ 
\midrule
Baseline with uniform random choice	 &	11371	& &	13.50 &	\\
\midrule
Minimizing interventions	&	11791 & ($+3.7\%$)	&	11.43 & ($-15.3\%$)	\\
Minimizing cost	            &	10768 & ($-5.3\%$)	&	17.43 & ($+29.1\%$)	\\
Agent preference	        &	11493 & ($-1.1\%$)	&	14.55 & ($+7.8\%$)\\
Pre-shielded agent	        &	6912 & ($-39.2\%$)	&	--    & --	\\
\bottomrule
\end{tabular}

%% file: conclusion.tex
\section{Conclusion}
\label{sec:conc}
We presented a practical approach to synthesize a near-optimal safety strategy via finite (2-player) abstractions of hybrid Markov decision processes, which are systems of complex probabilistic and hybrid nature.
In particular, we deploy a simulation-based technique for inferring the 2-player abstraction, from which a safety shield can then be constructed.
We show with high statistical confidence that the shields avoid unsafe outcomes in the case studies, and are significantly faster to construct than when deploying symbolic techniques for computing a correct 2-player abstraction.
In particular, our method demonstrates statistical safety on several case studies, two of which are industrial.
Furthermore, we study the difference between pre- and post-shielding, reward engineering and a post-shielding optimization.
In general, we observe that reward engineering is insufficient to enforce safety, and secondarily observe that pre-shielding provides better controller performance compared to post-shielding.

Future work includes applying the method to more complex systems, and using formal methods to verify the resulting safety strategies, maybe based on~\cite{ForetsFS20}.